# Physisorption of molecular oxygen on single-wall carbon nanotube bundles and graphite

Hendrik Ulbricht, Gunnar Moos, and Tobias Hertel
*Fritz-Haber-Institut der Max-Planck-Gesellschaft, Faradayweg 4-6, D-14195 Berlin, Germany*

We present a study on the kinetics of oxygen adsorption and desorption from single-wall carbon nanotube (SWNT) and highly oriented pyrolytic graphite (HOPG) samples. Thermal desorption spectra for SWNT samples show a broad desorption feature peaked at 62 K which is shifted to significantly higher temperature than the low-coverage desorption feature on HOPG. The low-coverage $O_2$ binding energy on SWNT bundles, 18.5 kJ/mol, is 55% higher than that for adsorption on HOPG, 12.0 kJ/mol. In combination with molecular mechanics calculations we show that the observed binding energies for both systems can be attributed to van der Waals interactions, *i.e.* physisorption. The experiments provide no evidence for a more strongly bound chemisorbed species or for dissociative oxygen adsorption.

## I. INTRODUCTION

Charge transfer by adsorbed oxygen is currently discussed as a potential source of doping for single-wall carbon nanotubes (SWNT). The *p*-doping of SWNT based field effect transistors [1,2] as well as the positive thermopower of SWNT samples [3,4] gave rise to speculations regarding possible doping by atmospheric gases including oxygen. Evidence for $O_2$ induced doping is provided by recent studies which find that the exposure of SWNT samples and devices to $O_2$ appears to have a strong influence on their electronic transport properties [5]. Chemical doping requires at least a weak chemical bond to facilitate charge-transfer from the substrate to the lowest unoccupied orbital of adsorbed oxygen. *Ab initio* calculations of $O_2$ interactions with different tube types performed within the local density approximation (LDA) yield binding energies ranging from 3.7 kJ/mol to 24.1 kJ/mol [6,7,8] with a charge transfer from the tube to the $O_2$ molecule of up to 0.1 $e^-$ which indeed seems to support oxygen induced doping. The adsorption of $O_2$ on graphite, however, is known to be entirely due to weak van der Waals forces which do not give rise to noticeable adsorbate-substrate charge transfer. The stretch frequency of $O_2$ adsorbed on highly oriented pyrolytic graphite (HOPG), for example, is unchanged with respect to the gas phase value [9]. Core level photoemission spectra likewise show no chemical shift or signs of charge transfer [10] and gradient-corrected LDA calculations do not provide any evidence for oxygen chemisorption on graphite [11].

It is, therefore, still discussed controversially what microscopic processes lead to the observed changes in the electronic transport properties of SWNTs. Apart from the aforementioned chemical doping [3,4,5,12] it has also been proposed that doping of the tube-metal contacts is responsible for changes in the transport properties [1,13,14]. It also remains unclear whether atomic or molecular oxygen is responsible for the charge transfer that is necessary for doping – let it be at contacts or on the tube surfaces. Here, the oxygen-SWNT binding energy may provide important information to elucidate the actual doping mechanism. The difficulties of *ab initio* calculations with the prediction of $O_2$-SWNT binding energies – which likely arise from the over-binding of LDA as well as the improper treatment of dispersion forces – clearly call for a complementary experimental investigation.

Here, we present a study on the kinetics of oxygen desorption and adsorption from SWNT samples and highly oriented pyrolytic graphite (HOPG). The experimental binding energy is compared to molecular mechanics calculations in trying to determine whether oxygen forms a weak chemical bond or if the interaction is purely van der Waals type – in which case chemical doping is not possible. We also studied if oxygen adsorbs dissociatively on the SWNT samples.

## II. EXPERIMENT

The SWNT samples, made from a commercially available nanotube suspension (tubes@rice, Houston, Texas) containing SWNTs with a diameter distribution peaked at 12 Å, are fabricated according to the procedure described elsewhere [15]. Samples were outgassed by repeated heating and annealing cycles under ultra high vacuum (UHV) conditions with peak temperatures of 1200 K. This procedure ensured that traces of solvent, carboxylic groups, or other functional groups left from the purification procedure were removed prior to adsorption experiments. UHV with a base pressure of $1 \cdot 10^{-10}$ mbar was maintained by membrane, turbo-drag and turbo-molecular pumps. The SWNT-paper was attached to a tantalum disk (1 cm diameter) by adhesive forces after wetting the sample and substrate with a droplet of ethanol. A sample of highly oriented pyrolytic graphite (HOPG) was mounted for reference and calibration experiments on the backside of the Ta-disk. A type-K thermolcouple, spot-welded to the Ta-disk, was calibrated for temperature control using the evaporation of $O_2$ multilayers in combination with the heat of oxygen sublimation, 9.2 kJ/mol [16]. The remaining uncertainty of the temperature measurement is estimated to be ±2 K. The sample holder was attached to a He-continuous flow cryostat, which enabled sample cooling down to 28 K. Molecular oxygen of 99.998% purity was admitted to the chamber through a retractable pinhole-doser with a 10 µm diameter orifice. The gas flux onto the sample was on the order of $10^{-11}$ mol $s^{-1}$ $cm^{-2}$ equivalent to about 0.01 ML $s^{-1}$. The monolayer (ML) coverage on HOPG corresponds to a close packed oxygen layer (upright $O_2$ molecules) with a nearest neighbor distance of 3.22 Å resulting in a density of 1.85 nmol $cm^{-2}$ [17]. Calibration of absolute coverages and desorption rates was achieved using this value in combination with the integral of the monolayer TD spectrum extending to 32 K (see Fig. 1). Desorbing oxygen was detected by a quadrupole mass spectrometer (Spectra Satellite) and the sample temperature was ramped at a constant rate (Lakeshore 340) of 0.25-0.50 K $s^{-1}$. Sticking coefficients were measured using a modified King and Wells-type setup [18] with a pinhole doser replacing the molecular beam apparatus. Additional experimental details have been published elsewhere [19,20].

## III. RESULTS AND DISCUSSION

### A. Kinetics of $O_2$ desorption

The interaction of $O_2$ with graphite has been studied extensively by a variety of techniques [9,10,17,21]. The $O_2$-graphite system with its rich phase-diagram can roughly be divided into low coverage δ, and high coverage ζ phases [21]. Below 27 K, the molecular orientation in the low coverage δ phase is parallel to the surface with a nearly rectangular oblique unit cell. At temperatures above about 27 K this phase undergoes a transition to a liquid phase where the $O_2$ molecular axis is not clearly oriented and the average nearest neighbor spacing is 4.0 Å [21]. Steric considerations suggest that molecules in the liquid phase undergo hindered rotations with some tilting away from the surface normal. During thermal desorption around 41 K, the adlayer therefore consists of a 2D dilute gas coexisting with 2D liquid islands.

Thermal desorption spectra of $O_2$ from HOPG after adsorption of 0.11-3.7 ML oxygen, at 29 K are shown in Fig. 1. The heating rate was 0.25 K $s^{-1}$. The desorption traces show high- and low-temperature features attributed to desorption from the first, second and higher monolayers. The change of the surface coverage Θ is given by the rate equation

$$\frac{d\Theta}{dt} = -\boldsymbol{n}\,\Theta^n \exp\left(-\frac{E_B}{k_B T}\right) \quad (1)$$

$n$ is the order of desorption, $\boldsymbol{n}$ the pre-exponential frequency-factor and $E_B$ is the binding energy. The shape of the sub-monolayer TD spectra with their exponentially increas-ing leading (low-temperature) edge and the sudden high-temperature cut-off is typical for desorption of inert, i.e. weakly interacting, gases from graphite surfaces [22,23]. The order of desorption, $n=0$, is attributed to the equilibrium of adsorbates desorbing from a 2D dilute gas on free patches of the surface with adsorbates leaving the surface from the second layer on top of 2D liquid or solid islands [22].

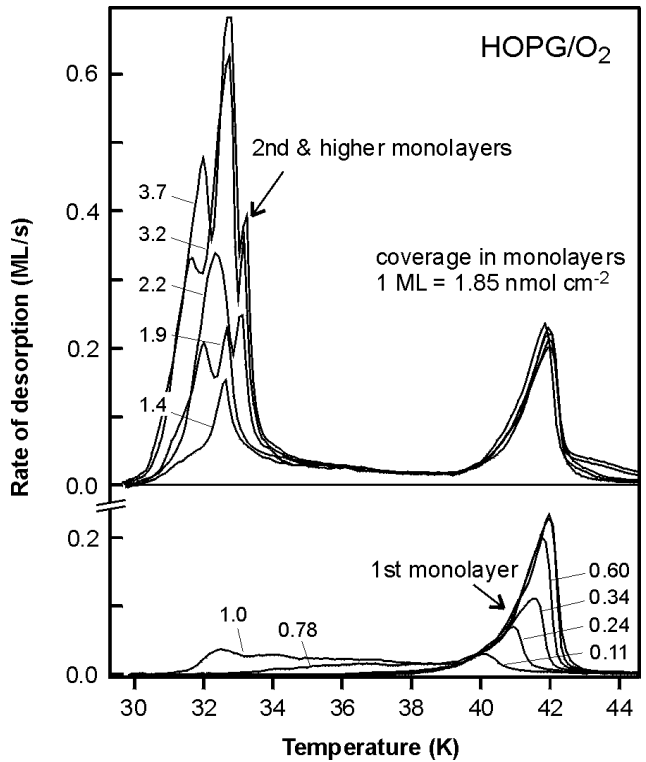

FIG. 1. Thermal desorption of $O_2$ from highly oriented pyrolytic graphite. The heating rate was 0.25 K $s^{-1}$.

Using a pre-exponential factor of $10^{15}$ $s^{-1}$ – as for the desorption from $O_2$ multilayers [16] – we obtain a binding energy of 12.3 kJ/mol from the high-temperature cutoff. Using the pre-exponential of $1.4 \cdot 10^{14}$ $s^{-1}$ obtained in the following section from adsorption experiments, we obtain a slightly lower binding energy of only 11.8 kJ/mol.

As seen in Fig. 2 the broad desorption maxima of TD traces from SWNT samples are systematically shifted to higher temperatures with respect to the monolayer desorption feature on HOPG. In addition they show no saturation at higher oxygen coverages which can be attributed to diffusion of $O_2$ from the sample surface into the bulk of the material. The TD spectra also do not show any trace of multilayer desorption up to coverages exceeding ~100 nmol·$cm^{-2}$ (here the area refers to the geometric, i.e. visible, area and not the specific sample area per unit mass). As seen from the inset in Fig. 2, the maximum of desorption traces slowly shifts to somewhat lower temperatures at coverages exceeding a few tens of nmol·$cm^{-2}$.

For the analysis of TD spectra from porous samples we have previously introduced a coupled desorption-diffusion

(CDD) model which can account for all of the experimental observations as discussed in detail for the Xe-SWNT system [20]. Within this model the rate of desorption is calculated by accounting for both, the desorption from the visible sample surface as given by equation (1) and the diffusion of adsorbates through the porous network of the sample. The latter is determined by the gradient of the concentration $C_g$, where $C_g$ refers to the amount of gas phase species in thermal equilibrium with adsorbates on the internal surfaces of the SWNT sample. The rate can thus be written as the sum of the desorption and bulk diffusion components which are both evaluated at $z=0$, *i.e.* at the surface of the SWNT sample,

$$j(0) = \frac{\tilde{f}}{N_A s} \frac{d\Theta(0)}{dt} + D_g \frac{\partial C_g(z)}{\partial z}\bigg|_{z=0} \quad (3)$$

Here $\tilde{f}$ is a factor close to 1 which corresponds to the increase in the visible sample surface due to microscopic surface roughness, $s$ is the adsorbate cross sectional area, $N_A$ is Avogadros number, and $D_g$ is a diffusion constant. It has been shown previously [20], that the change of the total adsorbate concentration $C=C_g+C_s$ within the sample ($C_s$ is the concentration of species adsorbed on the internal surfaces) and the resulting mass transport can be obtained from the simple one-dimensional diffusion equation:

$$\frac{\partial C(z)}{\partial t} = [D_g h(T) + D_s(1-h(T))]\frac{\partial^2 C(z)}{\partial z^2} \quad (4)$$

Here $h(T)$ is given by:

$$h(T) = \left(1 + \frac{k_B T b A_s}{f s}\right)^{-1} \quad (5)$$

with

$$b = \frac{s}{n\sqrt{2\pi m k_B T}} \exp\left(\frac{E_B}{k_B T}\right) \quad (6)$$

Equation (4) is a linear diffusion equation that can be solved numerically using the diffusion constants given in ref. [20]. As discussed previously we expect that the high effective diffusion barriers for migration along the surface of SWNT bundles favors mass transport by diffusion of gas-phase species $C_g$ through the porous structure of the sample. Using the same parameters employed for the analysis of xenon desorption from SWNT samples [20] we here obtain an $O_2$ binding energy of 18 kJ/mol which is about 50% higher than that obtained for $O_2$ on HOPG. The uncertainty of this binding energy resulting from a lack of information on some of the parameters within the CDD model is estimated to be about 15%. Note, that desorption features are found to shift to slightly higher or lower temperature if different bucky-paper samples are compared.

In trying to determine whether a chemisorbed or even dissociatively adsorbed oxygen species exists on the SWNT samples we have monitored the rate of $O_2$ desorption together with the rate of CO and $CO_2$ production during heating of the sample up to about 1000 K. Oxygen TD spectra show no evidence of an additional desorption feature at temperatures beyond 100 K. The CO and $CO_2$ pressure is found to increase monotonically during heating of the sample which is due to desorption from the sample mount or filaments and cannot be attributed to desorption from the SWNT substrate. This is evident from the absence of any distinct TD features as well as the fact that similar spectra are obtained if no $O_2$ was previously admitted to the sample. The probability of oxygen chemisorption or dissociation is thus too low to be detected in these experiments. LDA calculations by Lamoen and Person for graphite [11] as well as the calculations by Jhi *et al.* for an (8,0) nanotube [6] likewise indicate that the barrier to dissociative oxygen adsorption is very high, probably exceeding 100 kJ/mol in qualitative agreement with the results presented here.

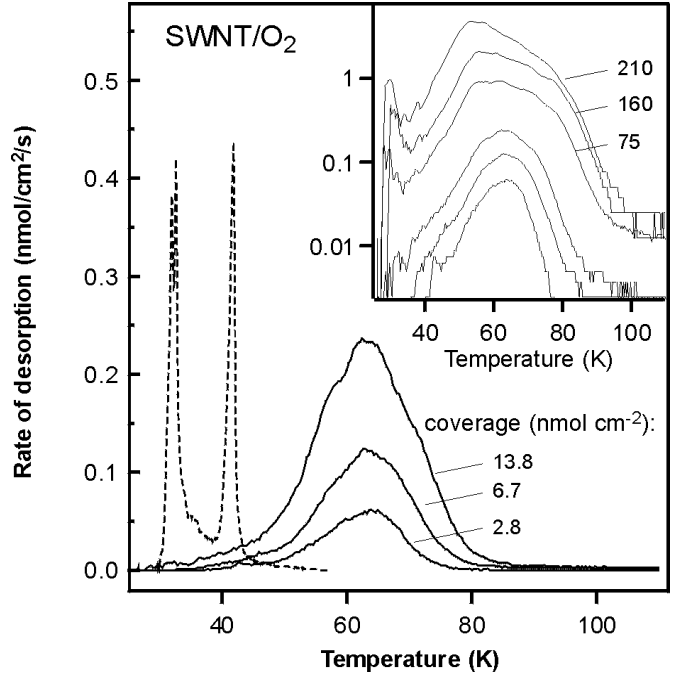

FIG. 2. Low and higher oxygen coverage (inset) thermal desorption traces from the SWNT sample and for comparison from HOPG (1.9 ML, dashed line). The heating rate was 0.5 K s$^{-1}$.

### B. Kinetics of $O_2$ adsorption

The rate of adsorption is generally characterized by the dimensionless sticking coefficient $s$ which is defined as the number of molecules trapped on the surface normalized to the incident flux $f$. For physisorption this coefficient is sometimes also referred to as condensation coefficient [18]. Experimentally, this can be obtained by a King and Wells type measurement [18] where the background pressure $P_1(t)$ in the chamber is recorded during exposure of the sample to a directed beam of gas and normalized using the pressure $P_2(t)$ for the same experimental run where the gas is not permitted to adsorb on the surface. Gas is prevented from adsorbing either by blocking the sample with an inert flag or by keeping it at elevated temperature. We define the adsorption probability $r$:

$$r = \frac{P_2 - P_1}{P_2} \quad (7)$$

If multiplied with the incident flux $f$ this gives the net change in surface coverage during an adsorption experiment:

$$rf = \frac{d\Theta}{dt} \quad (8)$$

The latter is equivalent to the balance between adsorption and desorption where $\frac{d\Theta}{dt}$ is given by

$$\frac{d\Theta}{dt} = sf - n\Theta^n \exp\left(-\frac{E_B}{k_B T}\right) \quad (9)$$

At low temperatures, where desorption is negligible, $r$ thus directly yields the sticking coefficient $s$. The uncertainty of the absolute values obtained for $s$ is estimated to be about 20%. Relative changes in $s$, however, have a smaller uncertainty of about ±0.05.

The probability $r$ for oxygen adsorption on SWNT and HOPG samples is shown in the inset of Fig. 3 as a function of surface coverage for a sample temperature of 34 K. The coverage is obtained by integration of $rf$ over time. Since desorption from both samples at low coverages and 34 K is negligible we can associate the initial probability with the low-coverage sticking coefficient $s_0$. Within experimental errors, the latter is found to be 0.9 for HOPG and about 1.0 for the SWNT sample. In the inset of Fig. 3 it can be seen that the sticking coefficient on HOPG increases by about 10% before it drops sharply to zero near completion of the monolayer due to increasing desorption from the second layer on which $O_2$ is less strongly bound. Due to diffusion of gas into the bulk of the material the behavior on the SWNT sample is found to be quite different. The adsorption probability of the SWNT sample stays nearly constant for an extended period of time with no significant decrease of the sticking probability up to up to an exposure of at least 100 nmol cm$^{-2}$.

The initial, low-coverage adsorption probability $r_0$ is plotted as a function of the sample temperature in Fig. 3. It is found to decrease rapidly once $T_{max}$, the temperature at which the desorption rate reaches its maximum, is exceeded. Due to the dependence of eq. (9) on the binding energy $E_B$ one can use these measurements as an alternative means to determine the oxygen binding energy. For simplicity we will assume that the low coverage sticking coefficient is independent of temperature over the temperature range of interest. This is reasonable as the temperature dependence of $s_0$ is generally small [18,24]. In this case the decrease of $r_0$ can be attributed exclusively to changes of the rate of desorption.

Assuming zero order desorption for the oxygen-graphite system we obtain a binding energy of 11.6 kJ/mol and a pre-exponential factor $1.4 \cdot 10^{14}$ s$^{-1}$ from a best fit to our experimental data (upper left part of Fig. 3, dashed line). The binding energy obtained from these adsorption measurements is thus in good agreement with that determined from TD spectra in the previous section and we will use an average over both experiments, 12.0±0.5 kJ/mol, as value for future reference. Previous studies quote a slightly smaller value for the isosteric heat of adsorption of 10.5 kJ/mol [25] and 10.0 kJ/mol [26]. This discrepancy probably arises from the different surface coverage at which the binding energy was evaluated. Our value refers to the low coverage binding energy in the liquid phase. Note, that the isosteric heat of adsorption $q_{st}$ has to be corrected using $q_{st} = E_B + 2k_B T$ for a direct comparison with binding energies $E_B$.

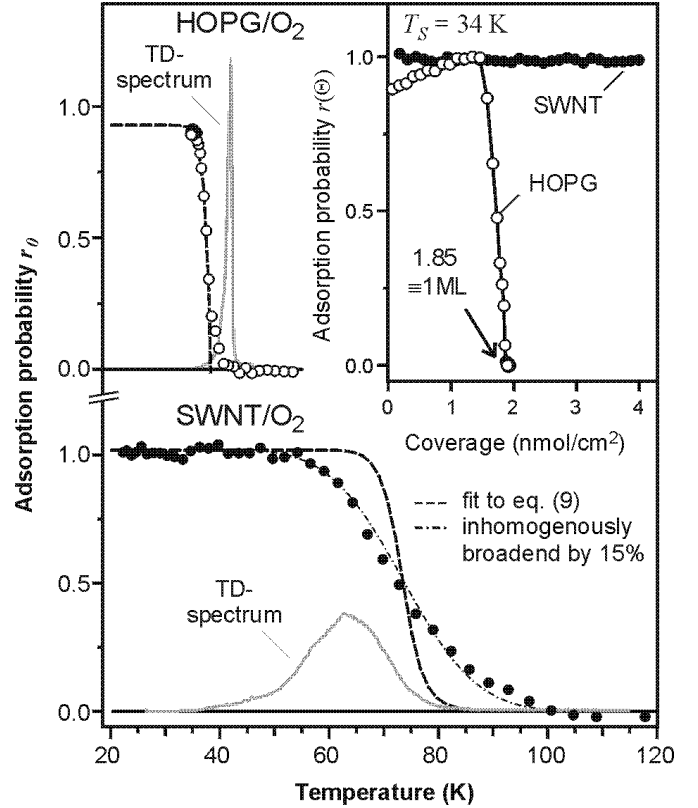

FIG. 3. Low coverage adsorption probability for HOPG (open circles) and SWNT samples (solid circles) as a function of the sample temperature and coverage (upper right).

A fit of eq. (9) to the temperature dependence of $r$ using first order kinetics for the SWNT sample is shown in Fig. 3. First order kinetics is appropriate for the SWNT sample where – in contrast to graphite – ordered adsorbate structures likely do not exist in the temperature range of interest. Here we assume that diffusion processes can be neglected in the zero coverage limit. The binding energy obtained from the fit in Fig. 3 using a pre-exponential factor of $10^{15}$ s$^{-1}$ is 19 kJ/mol in good agreement with the value obtained from the analysis of desorption traces, 18 kJ/mol. For future reference we will thus use 18.5 kJ/mol. It is clear, however, that the slope of the experimental trace near the inflection point is much smaller than that of the fit as given by eq. (9). We suggest that this is indicative of some inhomogeneous broadening due to a distribution of binding energies. Much better agreement with the experimental results is obtained if we allow for a Gaussian spread of binding energies $E_B$ around the mean value with $\Delta E_B = 3$ kJ/mol (dot-dashed line in Fig. 3).

### C. Molecular mechanics calculations

The binding energy for $O_2$ adsorbed on graphite is generally attributed to dispersion-, *i.e.* van der Waals forces. Electron energy loss studies, for example, reveal that the stretch frequency of adsorbed $O_2$ of 191 meV [9] is unchanged with respect to the gas phase value, 194 meV [27]. Core level photoelectron spectra likewise show no significant shift relative to gas phase values [10]. Any charge transfer leading to the formation of chemisorbed peroxo $O_2^-$ species would lead to a weakening of the $O_2$ bond and would be evident from a pronounced shift of the O1s core-level in photoemission as well as a reduction of the stretch frequency to about 140 meV [28]. The experimental findings are thus considered as fingerprints for physisorption and reflect the purely VDW type forces between $O_2$ and the graphite surface. *Ab initio* calculations including gradient corrections to the LDA functional for $O_2$ adsorption on graphite also do not provide evidence for a chemisorbed $O_2$ species [11].

It is therefore well justified to use VDW potentials for C-O and O-O interactions in the calculation of binding energies for physisorbed oxygen. Here, we use an empirical van der Waals potential of the form:

$$V(r) = C\exp(-\boldsymbol{a}r) - Dr^{-6} \qquad (10)$$

where $r$ is the distance between interacting carbon and oxygen atoms and $C$, $\boldsymbol{a}$, and $D$ are element specific constants. Such potentials have been used successfully to explain many of the structural properties and the rich phase diagram of the oxygen-graphite system [29,30,31]. Additional forces arising from electric multipole interactions, substrate mediated dispersion interactions or magnetic spin-spin interactions are generally assumed to be comparatively weak and do not contribute significantly to the total binding energy [30].

We begin with a comparison of experimental and calculated binding energies for graphite. The binding energy of $O_2$ to graphite is calculated for structural parameters characteristic of the high temperature liquid phase, from which $O_2$ desorbs at 42 K (nearest neighbour spacing 4.0 Å [21]). Using the ($C$, $\boldsymbol{a}$, $D$) parameters (1120.6 eV, 3.68 Å$^{-1}$, 12.91 eV Å$^6$) by Etters and Duparc for O-O interactions and an O-substrate interaction given by the VDW potential of Stele [30,32], we obtain an oxygen binding energy of 8.1 kJ/mol with a contribution of 35% from $O_2$-$O_2$ interactions within liquid islands. Somewhat better agreement with the experimental binding energy was obtained using VDW parameters of the MM3 force-field by Allinger [33]. Allingers VDW parameters ($C$, $\boldsymbol{a}$, $D$) for O-O interactions are (473.2 eV, 3.30 Å$^{-1}$, 13.46 eV Å$^6$) while C-C interactions are given by (449.2 eV, 3.06 Å$^{-1}$, 19.93 eV Å$^6$). C-O interactions can be obtained from the usual combining rules [33]. These parameters yield a binding energy of 10.3 kJ/mol with a 25% contribution from $O_2$-$O_2$ interactions. Here, the underestimation of experimental binding energies by about 20-30% presumably arises from the particular choice of VDW parameters.

In the following we will explore if the higher binding energy found experimentally for the SWNT/$O_2$ system can be accounted for by van der Waals interactions. To this end, binding energies are calculated for various molecular orientations and sites on a rigid, close packed (9,9)-SWNT bundle with a fixed lattice constant of 15.6 Å. The VDW parameters are those of Allingers MM3 force field. If the $O_2$-axis is aligned parallel to SWNTs we obtain binding energies of 14.9 kJ/mol, 14.3 kJ/mol and 15.4 kJ/mol for adsorption in the groove, interstitial channel and endohedral sites, respectively (see Fig. 4 and Table I). Due to the comparatively high zero point energies of molecular vibration and frustrated rotation within the interstitial channel, however, one has to correct the corresponding binding energy. The calculated vibrational quantum for frustrated rotation of $O_2$ is about 12 meV while that for vibration of the molecules center of mass is about 13 meV. This would lead to a reduction of the binding energy in the interstitial site of approximately 1.2 kJ/mol. Note, that binding energies for the interstitial channel are very sensitive to the repulsive part of the VDW potential and depend strongly on the bundle lattice constant, which is expected to give rise to considerable uncertainty.

Adsorption with the molecular axis aligned perpendicular to the bundle surface with a binding energy of 12.9 kJ/mol for the groove site and 12.7 kJ/mol for adsorption inside of tubes with the molecular axis perpendicular to the wall, is slightly less favorable. Steric constraints in the interstitial channel prohibit such a rotation.

At very low temperature the condensation in one-dimensional oxygen chains should give rise to small corrections to the binding energies calculated above. The possible alignment of the oxygen axis is a) parallel to the chain axis ( $\parallel$ , $\alpha$-chains) and b) perpendicular to the chain axis ($\perp$ , $\beta$-chains). For adsorption in the grooves on the external bundle surface the results in Table I indicate that the slightly higher energy for condensation in $\beta$-chains is probably not sufficient to overcome the energy gain of $O_2$-tube interactions when the molecular axis is parallel to the grooves. Oxygen molecules, therefore most likely orient themselves parallel to the tubes if adsorbed in groove sites at low temperature and coverage. At higher coverages steric constraints may induce a reorientation of the $O_2$ axis perpendicular to the chain in analogy to the low temperature, high coverage $\epsilon$ phase on graphite. At very low temperatures spin-spin interactions and magnetic ordering probably play an increasingly important role for the order in the adsorbate phase. We assume, however, that ordered one-dimensional phases are not stable up to desorption temperatures and that condensation does not need to be considered for a comparison with experimental binding energies. Note, that the 2D oxygen liquid phase coexists with the 2D gas on graphite only up to about 60 K [21].

TABLE. I. Summary of $O_2$ binding energies (in kJ/mol) obtained from the molecular mechanics calculations for SWNT bundles and graphite. The orientation of the $O_2$ symmetry axis was chosen to be either parallel ($\parallel$) or perpendicular to the tubes and the surface ($\perp$). For oxygen islands or chains the alignment of the $O_2$ axis is within ($\parallel$) or perpendicular ($\perp$) to the adsorbate layer.

| | $E_B$ (Expt) | $E_B$ (Theo.) |
|---|---|---|
| Graphite liquid phase | 12.0 | 8.1 [a,c] |
| | | 10.3 [b,c] |
| SWNT bundle | 18.5 | 14.9$^{\parallel}$ / 12.9$^{\perp}$ (groove site [b]) |
| | | 14.3$^{\parallel}$ / --- (interstitial site [b]) |
| | | 15.4$^{\parallel}$ / 12.7$^{\perp}$ (endohedral site [b]) |
| 1D $O_2$ chains [b] | - | 0.5$^{\parallel, \alpha\text{-chain}}$ / 1.1$^{\perp, \beta\text{-chain}}$ |
| 2D $O_2$ islands [b] | - | 3.3$^{\perp}$ |

[a] Parameters by Etters and Duparc [30]
[b] Parameters by Allinger [33]
[c] Including $O_2$-$O_2$ interactions

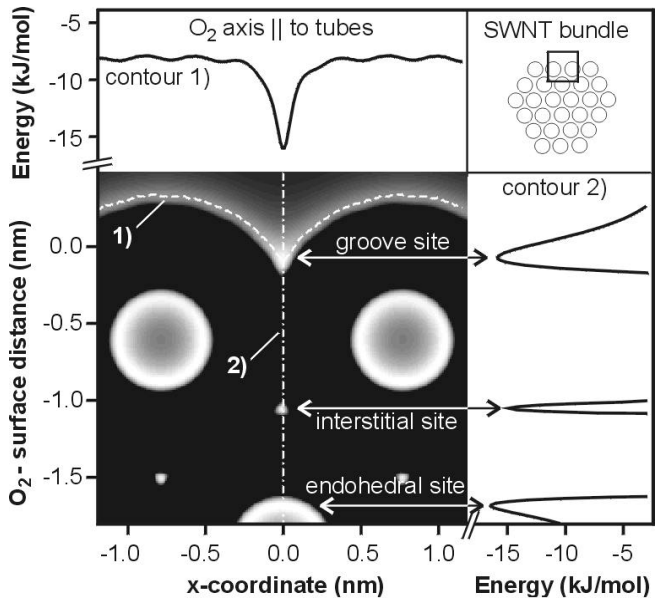

FIG. 4. Two-dimensional potential energy surface for oxygen with its molecular axis aligned parallel to the bundle and tubes axis. Bright areas correspond to regions with high binding energy.

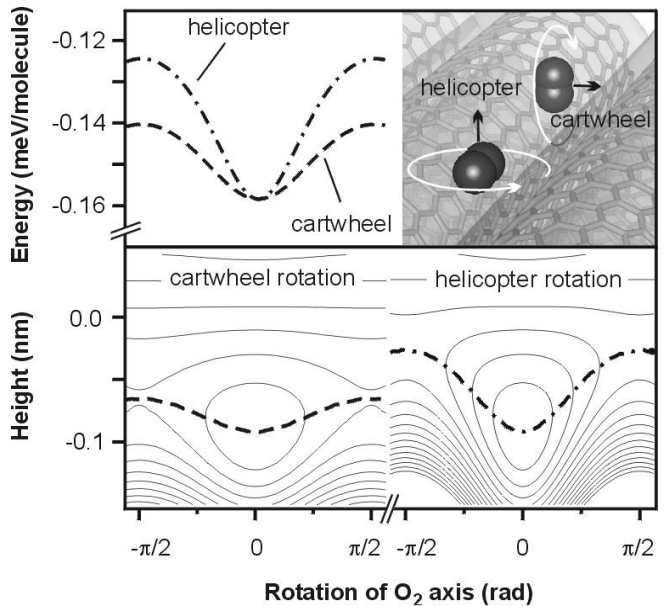

FIG. 5. Two-dimensional potential energy surfaces for oxygen molecules rotating in the groove sites on the external bundle surface. In contrast to flat surfaces the barrier to cartwheel-type rotations is found to be lower than that for helicopter-type rotations. The contour spacing is 10 meV.

As a last point of interest we will discuss the energetics calculated for the excitation of different molecular degrees of freedom. The translational degrees of freedom, for example, can be characterized by basically free motion of adsorbed molecules inside of tubes and parallel to the bundle axis but high barriers for translation perpendicular to the bundle axis on the external bundle surface. The migration along the bundle perimeter, for example, is associated with a 7 kJ/mol barrier – approximately half the binding energy in the external groove (see Fig. 4). A similar ratio of this migration barrier to the binding energy in groove sites is found for xenon [20]. This deep potential well leads to lateral strongly confined adsorbate phases on the external bundle surface which should be stable up to the temperature of desorption. In contrast, rotational degrees of freedom in the groove site are expected to be strongly excited at the temperature of desorption. As seen from Fig. 5, rotational melting is expected to proceed via the excitation of cartwheel like rotations with a barrier to free rotation of about 2 kJ/mol and then by helicopter type rotation with a barrier of over 3 kJ/mol. This is in contrast to the behavior on flat surfaces where the helicopter mode is generally much lower in energy.

Summing up the results of this section: oxygen binding energies are calculated for various sites on an idealized SWNT bundle using van der Waals potentials. If compared with energies calculated for oxygen binding to graphite we find a 30-50% increase for different sites on the SWNT bundles. This is in good agreement with the experimentally observed increase from 12.0 kJ/mol to 18.5 kJ/mol, by about 55% and suggests that binding of $O_2$ to SWNT samples is – as for graphite – entirely through dispersion forces.

## VI. CONCLUSIONS

We have studied the kinetics of oxygen adsorption and desorption from HOPG and SWNT samples. The experimental low coverage binding energy for adsorption on HOPG, 12.0±0.5 KJ/mol increases to 18.5 kJ/mol by about 55% when oxygen is adsorbed on SWNT samples. Molecular mechanics calculations of the oxygen-SWNT bundle interaction using van der Waals potentials show that the observed change of the binding energy can be attributed entirely to an increase of the effective coordination in binding sites on SWNT bundles which is characteristic of physisorption.

Using the experimental binding energy in combination with simple Langmuir kinetics we estimate that the equilibrium room temperature oxygen coverage on SWNT samples at atmospheric conditions is of the order of $10^{-5}$ molecules per carbon atom. It thus seems unlikely that the observed *p*-doping of un-annealed SWNT samples or the influence of oxygen on the electronic transport properties of SWNT based devices is due to charge transfer by such weakly bound oxygen species. This and the fact that we find no evidence for a chemisorbed species suggests that the observed *p*-doping must be due to a minority oxygen species. This minority species may be located either at defect sites on the SWNT bundles as suggested previously or at tube-metal contacts in electronic devices [1,14]. The time constants, characteristic of the changes following degassing or oxygen doping of SWNT samples at high temperatures and pressure, ranging from seconds to minutes [4,5], are likewise indicative of a strongly bound minority species that is responsible for changes in the conductivity of SWNT samples.

Our experimental results thus strongly suggest that oxygen binds to these SWNT samples, through dispersion forces and not by the formation of a chemical bond. We conclude that the transport properties of SWNT samples are not influenced by charge doping from this physisorbed oxygen species.

## ACKNOWLEDGEMENTS

We thank G. Ertl for his continuing support of this work.